\documentstyle[twoside,fleqn,espcrc2,epsf]{article}
\newcommand{\bi}{\begin{itemize}}
\newcommand{\ei}{\end{itemize}}
\newcommand{\beq}{\begin{equation}}
\newcommand{\eeq}{\end{equation}}

\newcommand{\fig}[1]{Fig.~\ref{#1}}
\newcommand{\tab}[1]{Tab.~\ref{#1}}
\title{Hybrid and Orbitally Excited Mesons in Full QCD \thanks{
    Presented by P. Lacock}}
\author{ 
        P. ~Lacock$^{\rm a}$ and K. Schilling $^{\rm a,b}$ \\  
        (SESAM Collaboration) \\ [8pt] 
{\rm $^a$}HLRZ, c/o Forschungszentrum J\"ulich, D-52425 J\"ulich,
          and DESY, D-22603 Hamburg, Germany       
{\rm $^b$}Physics Department, University of Wuppertal,
           D-42097 Wuppertal, Germany\\[8pt]}
\begin{document}
\begin{abstract}
 
We present results for 
the hybrid meson spectrum produced by gluonic excitations in full QCD 
using Wilson fermions.
For  the  spin-exotic
mesons with $J^{PC}=1^{-+},\ 0^{+-}$, and $2^{+-}$ 
we find the lightest state to be $1^{-+}$ with
a mass of 1.9(2) GeV.
Results obtained for orbitally excited mesons are also presented. 
\end{abstract}
\maketitle


\section{INTRODUCTION}

A quantitative study of the
QCD hadronic spectrum should
include a study of states with excited orbital angular momentum.
These states generally require non-local operators in order to construct
observables  
with the correct symmetries.


Another topic of interest that goes  
beyond the determination of the ground state spectrum is a study of 
gluonic mesons:  the glueballs and hybrid mesons.  Of
special interest are the hybrid meson states with $J^{PC}$ quantum
numbers that are not allowed in the quark model, the so-called
$exotics$. These include the $J^{PC}$ values $1^{-+},\ 0^{+-}$, and
$2^{+-}$.

In the quenched theory, the above mentioned topics have been 
investigated by two separate groups \cite{hyb,milchyb}. Here
we report on an initial study of hybrid and $L-$excited mesons
in two flavour QCD. The aim of the study
is to study the effects of dynamical quark effects. A separate
study by the MILC Collaboration using improved actions is 
also underway \cite{milchyb}.

\section{LATTICE OPERATORS}


In order to construct lattice operators with the desired angular
momentum or gluonic excitation, we have to combine representations of
the `spin' cubic group (coming from the quark spinors) with the
`orbital' cubic group (coming from the spatial paths). Following
\cite{hyb} we study non-local gluonic fields in specific
representations of the $lattice$ rotation group.


We work on a $16^3 \times 32$ lattice at $\beta$ = 5.6 
using 2 degenerate Wilson fermions.  
We use 4 $\kappa_{sea}$ values:
0.1560, 0.1565, 0.1570 and 0.1575, which correspond to 
$m_{\pi}$/$m_{\rho}$ = 0.83, 0.81, 0.76, 0.69 respectively.
At each $\kappa_{sea}$
value we have around 200 independent configurations
generated as part of the SESAM Collaboration endeavours. 
To improve statistics we additionally 
use a second time source at $t=$11.

At each $\kappa_{sea}$ value we generate a set of local propagators 
consisting of sources for the quark at $(1,1,1,t)$ and at $(1,1,7,t)$ for
the anti-quark, where $t$ =1 and 11.  
 At the source the propagators are connected by a path
$\cal P$ consisting of  fuzzed gluon links. The path $\cal P$ is chosen
to provide the desired angular momentum (in the case of the
$L$-excited mesons) or the gluon excitation (for the hybrid mesons). The
resulting hadronic correlations are therefore also by
definition gauge invariant. Fuzzed links are used to improve the overlap
with the ground state \cite{hyb}.

An alternative procedure to the method used here is to
study hybrid mesons using the continuum symmetries \cite{milchyb}.

\begin{figure}[htb]
\vspace{-0.9cm}
\epsfxsize=7.5cm\epsfbox{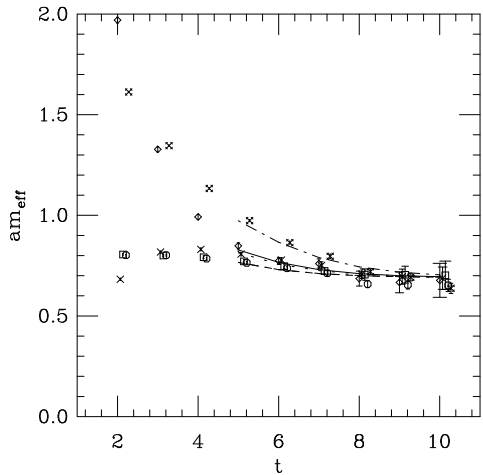}
\vskip -1cm
\hskip -6mm
 \caption{ The lattice effective mass for the $1^{++}$ meson vs. time $t$.
 For a discussion of the different operators used, see ref. \cite{hyb}.
\label{A1}}
\vspace{-0.9cm}
\end{figure}

\vspace{-0.3cm}
\section{$L$-EXCITED MESONS}

$L$-excited mesons can be studied by choosing $\cal P$ to be the
straight product of  fuzzed links connecting the quark and anti-quark at
the source and sink. At the source both the direction (here the \^z or 3
direction) and the length (=6) are fixed. At the sink, on the other
hand, we have three possible spatial directions, and the length $R$ can
also be varied.

We perform correlated 2-state fits to the effective mass and use as many
operators (i.e. different spatial link combinations and  choices of $R$)
as possible
to constrain the fits as much as possible. As a typical example we show
the results for the  $1^{++}$ meson in \fig{A1}.

We are also able to determine the hyperfine splitting among the four
states in the $P-$wave multiplet, finding that  the singlet state has
the lowest mass, while the other three members have masses which are
degenerate within the statistical errors. Similar behaviour was
observed in the quenched theory \cite{hyb}.

\section{HYBRID MESONS}

Hybrid mesons are by definition mesons with an excited gluonic
component.  From studies of static quarks it was found that the lowest
lying  hybrid states have  colour flux from the quark to
anti-quark excited in a transverse spatial plane~\cite{cmper}.  This can
be achieved by the choice of U-shaped paths of fuzzed links at the
source and sink. These operators were also successfully used
in a study of lighter quark masses in quenched QCD \cite{hyb}.

The lowest lying gluonic excitations have  spatial symmetries
corresponding to $L^{PC}$ = $1^{+-}$ and $L^{PC}$ = $1^{-+}$. Combining
these  with the $q \bar q$ spin representations, we obtain a range
of possible $J^{PC}$ values which include the spin-exotic quantum
numbers  $J^{PC}=1^{-+},\ 0^{+-}$, and $2^{+-}$ which are not present 
in the quark model. 

\begin{figure}[htb]
\vspace{-1.0cm}
\leavevmode
\epsfxsize=7.5cm\epsfbox{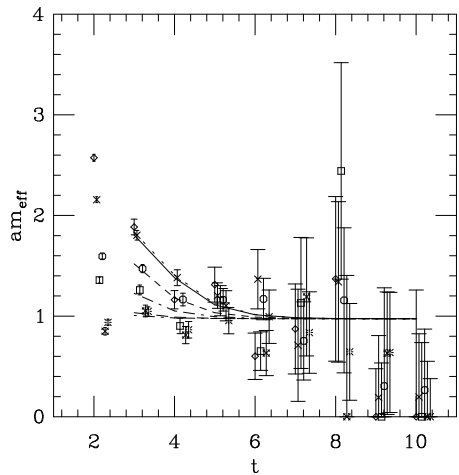}
\vskip -1cm
 \caption{ The lattice effective mass for the $J^{PC}=1^{-+}$  exotic
hybrid meson vs $t$. The sources used are U-shaped paths 
as well as closed colour loops, while at the sink we use only
 U-shaped paths.
 \label{exot1}}
\vspace{-8mm}
\end{figure}

In our simulation the source necessarily again has longitudinal length
($r$) = 6 fixed, while the transverse length ($d$) is free. At the sink
one can vary both ($R$ and $D$ respectively). From earlier experience we found
that the optimum  choice of operators has ($r,d$)=(6,6) at the source,
while at the sink we use ($R,D$) = (1,1), (3,3) and (6,6). The choice of
(6,6) at both source and sink moreover gives an upper bound on the
ground state mass.

 
\begin{table}
\vspace{-1mm}
\footnotesize
\begin{tabular}{llll}
multiplet & $J^{PC}$ & ma  (chiral) & mass (GeV)  \\
\hline
$\hat \rho$ & $1^{-+}$ $3^{-+}$ & 0.822 (100) & 1.9 (2) \\
$\hat a_0$ & $0^{+-}$ $4^{+-}$ & 1.015 (250) & 2.3 (6) \\
$\hat a_2$ & $2^{+-}$ $4^{+-}$ & 0.867 (500) & 2.0 (1.1) \\
\hline
\vspace{8pt}
\end{tabular}
\caption{The (physical) masses of the exotic mesons.}
 \label{RES}
\vspace{-8mm}
\end{table}

\begin{figure}[htb]
\vspace{-0.9cm}
\leavevmode
\epsfxsize=7.5cm\epsfbox{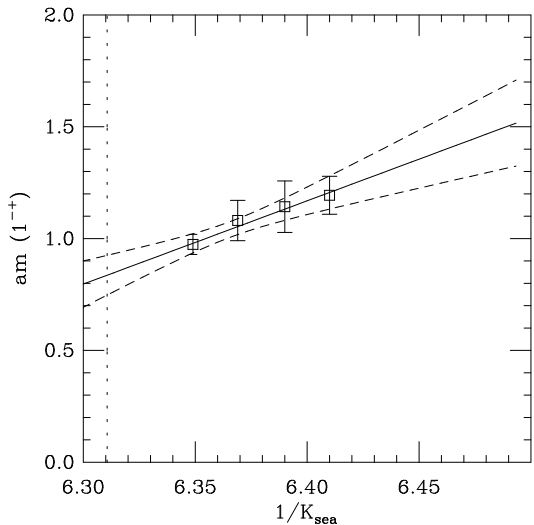}
\vskip -1cm
\caption{
 Extrapolation of the $1^{-+}$ exotic state to the chiral limit.
The vertical dashed line is given by
 $(\kappa_{sea}^{light})^{-1} \simeq$ 6.31066  \cite{sesam}.
 \label{exot1ch}}
\vspace{-8mm}
\end{figure}

The result for the effective mass of the $1^{-+}$ exotic meson is 
shown in \fig{exot1}, while in \fig{exot1ch} we show the extrapolation
for the same state to the chiral limit. The value thus obtained, 
together with those for the other 2 exotic states under consideration
here, are listed in \tab{RES}. The $1^{-+}$ exotic is again found
to be the lightest state.
This result is also found by the MILC investigation \cite{milchyb}.
 In order to convert the results into 
physical units we use $a^{-1} = 2.30$ GeV  determined
from $m_{\rho}$ \cite{sesam}. These values are given in the last 
column in \tab{RES}.

The estimate for the 
$1^{-+}$ state agrees -- within the given statistical
uncertainty -- with the value obtained in the quenched theory. 
One reason for this might be that the lattice volume 
($\approx$ 1.4 fm) is too small -- the wave function of
hybrid mesons is expected to be large. 
 To investigate these 
finite volume effects a
calculation of the hybrid spectrum on a
$24^3 \times 40$ lattice is currently under way.

Taking into account the statistical uncertainty
and likely finite volume effects, our result
is in fairly good agreement with
the 1.6 GeV candidate for the $1^{-+}$  state proposed in \cite{page}. 

\begin{figure}[htb]
\vspace{-1.4cm}
\leavevmode
\epsfxsize=7.5cm\epsfbox{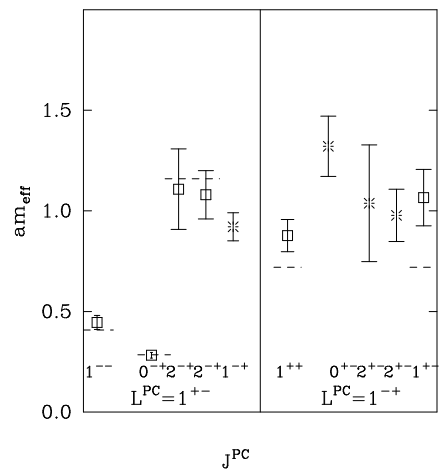}
\vskip -1cm
\caption{
 Ordering of the hybrid meson levels. The bursts denote the $J^{PC}$
exotic states.
 \label{order}}
\vspace{-6mm}
\end{figure}

In \fig{order} we show the ordering of the hybrid meson levels. The
dashed lines represent $L$-excited states.
The strong mixing of the
hybrid mesons and standard mesons (i.e. with no gluonic excitation) with
the same $J^{PC}$ values, 
already seen in the quenched approximation, is also apparent here.
In full QCD there are also other mixing effects: even exotic 
mesons can now mix with 4q ($q \bar q q \bar q$) states 
(e.g. $\pi \eta$). These effects are currently under investigation.
We are also planning to implement a third time source to 
reduce the statistical errors of the results presented above.

We acknowledge help from our colleagues from the SESAM Collaboration,
and would like to thank Mike Peardon for many useful discussions. 


\vspace{-5mm}
\begin{thebibliography}{99}
\frenchspacing
%


\bibitem{hyb}   UKQCD Collaboration, P. Lacock et al., 
Phys. Rev. D {\bf 54} (1996) 6997; Phys. Lett B {\bf 401} (1997) 308;
Nucl. Phys. B (Proc. Suppl.) 63A-C (1998) 203.

\bibitem{milchyb} C. Bernard et al., 
Nucl. Phys. B (Proc. Suppl.) 63A-C (1998) 206;
Phys. Rev. D {\bf 56} (1997) 7039;
and these proceedings
(presented by C. McNeile).
 
\bibitem{cmper}S.J. Perantonis and C. Michael, Nucl. Phys. B {\bf 347} (1990) 854.
 
\bibitem{sesam}N. Eicker et al., (hep-lat/9806027).
 
\bibitem{page} P. Page, LANL Preprint
(hep-ph/9806233).
 
 
 
%
\end{thebibliography}
\end{document}